\documentclass[a4paper]{jpconf}
\input epsf
\def\be{\begin{equation}}
\def\ee{\end{equation}}
\def\bea{\begin{eqnarray}}
\def\eea{\end{eqnarray}}

\def\veck{\underline{k}}

\def\momvol{\frac{d^{3}\underline{k}}{(2\pi)^{3}}}
\def\nn{\nonumber}
\newcommand{\lrb}{\left(}
\newcommand{\rrb}{\right)}
\newcommand{\ba}{\begin{array}}
\newcommand{\ea}{\end{array}}

\def\adag-k{a^{\dag}_{-k}}
\def\a-k{a_{-k}}
\def\ak{a_{k}}

\def\ddag-k{d^{\dag}_{-k}}
\def\d-k{d_{-k}}
\def\dk{d_{k}}

\def\Bdag-k{A^{\dag}_{\theta(k)}}
\def\B-k{A_{\theta(-k)}}
\def\Bk{A_{\theta(k)}}

\def\Ck{D_{\theta(k)}}

\def\C-k{D_{\theta(-k)}}
\def\Cdag-k{D^{\dag}_{\theta(k)}}

\def\Fk{F_{k}}
\def\Fdagk{F^{\dag}_{k}}
\def\F-k{F_{k}}
\def\Fdag-k{F^{\dag}_{-k}}

\def\Adag-k{A^{\dag}_{-k}}
\def\A-k{A_{-k}}

\def\Ddag-k{D^{\dag}_{-k}}
\def\Gdagk{G^{\dag}_{k}}
\def\Gdag-k{G^{\dag}_{-k}}
\def\G-k{G_{-k}}
\def\Gk{G_{k}}

\def\bmat{\lrb\begin{array}{c}}
\def\emat{\end{array}\rrb}

\begin{document}
\title{Pions emerging from an Arbitrarily Disoriented Chiral Condensate.}
\author{{\bf B.Bambah$^1$}and\\ {\bf K.V.S. Shiv Chaitanya}}
\address{$^1$ School of Physics \\ University of
  Hyderabad, Hyderabad-500 046,India}
\ead{bbsp@uohyd.ernet.in}
\author{\bf C.Mukku$^2$}
\address{$^2$ Department of Mathematics Panjab University, Chandigarh-160014,India}
\address{$^2$ Department of Computer Science \& Applications\\ Panjab University, Chandigarh-160014,India}
\ead{c\_mukku@yahoo.com}
Relativistic heavy ion collisions either with fixed target or with
head on collisions are providing results which suggest that the
chiral phase transition does take place with temperatures crossing
170Mev. The comparatively large sizes of the heavy ions involved
in the collisions along with their energy/nucleon, provides a
large interaction region for the quarks and gluons. This region is
where the Quark Gluon Plasma (QGP) forms. The picture we therefore
have is that of a collision region where a QGP is formed in a
highly non-equilibrium situation with chiral symmetry being
restored. This plasma then expands and cools, causing the restored
chiral symmetry to be broken once again. In a recent paper [1], we
have given a quantum field theoretical model providing
multiplicities of pions which would be observed if the QGP
undergoes a rapid quench in its expansion or if the expansion is
adiabatic. It was also shown how the rapid quench scenario, which
is appropriate for the formation of the disoriented chiral
condensate (DCC) has a enhancement in the multiplicities of the
observed pions. Thus the enhancement signatures in the pion
multiplicities are signals for the formation of the DCC also. In
our analysis, we had assumed the expansion of the QGP to be
isotropic for simplicity. However, recent results from RHIC
suggest that there may be anisotropy built in from the nature of
the collisions. Not all collisions will be head on with zero
impact parameter. There will be instances where there will be
non-zero impact parameter causing the impact region to be
elliptical and the corresponding QGP to have an elliptical flow.
From the nature of our quantum field theoretical model, borrowing
from the studies of scalar fields on expanding universes, such as
the Friedmann-Robertson-Walker universe, it is easy to extend the
study to include anisotropy such as that caused by an elliptic
flow. We shall report these results elsewhere, in a separate
communication [2]. Here, we shall be examining another aspect of
the restoration of chiral symmetry. The particular form of our
model allows for many metastable states through which the DCC can
form. In particular, we find that mixing between various erstwhile
pion states in the collision region can produce different signals
in the multiplicities of the final state pions.\\ We start with an
$O(4)$ quartet of scalar fields in order that we can construct the
dynamics of quenched pions in the formation of a disoriented
chiral condensate. The background field is the classical
disoriented vacuum and the Hamiltonian of the quantum fluctuations
around the background field is derived from the $O(4)$ sigma
 model keeping one-loop quantum corrections (quadratic order in the fluctuations)[1].
The resulting dynamical Hamiltonian for the pion and sigma fields
fields in terms of the neutral , charged  pion and sigma creation
and annihilation operators ($a,a^{\dag},c,c^{\dag},b $ $b^{\dag}$,
and $d,d^{\dag}$)  is[1]: \bea
 H&=&\int \momvol
\frac{1}{2}\{
\frac{\omega_{\pi}}{a^3}(a_k^{\dag}a_k+a_ka^{\dag}_k)
+\frac{\omega_{\pi}}{2a^3}(\frac{\Omega_{\pi}^2}{\omega_{\pi}^2}-1)(a_k^{\dag}a_k+a_ka_k^{\dag}+a_{-k}a_{k}+a_{-k}^{\dag}a_{k}^{\dag})\\\nn
\\ \nn
&+&\frac{\omega_{\Sigma}}{a^3}(
 d_k^{\dag}d_k+d_kd^{\dag}_k)+\frac{\omega_{\Sigma}}{2a^3}(\frac{\Omega_\Sigma^2}{\omega_{\Sigma}^2}-1)(d_k^{\dag}d_k+d_kd_k^{\dag}+d_{-k}d_{k}+d_{-k}^{\dag}d_{k}^{\dag})\}\\
 \nn &+&
 \frac{\omega_{\pi}}{a^3}(b_k^{\dag}b_k+c_kc^{\dag}_k)
+\frac{\omega_{\pi}}{2a^3}(\frac{\Omega_{\pi_\pm}^2}{\omega_{\pi}^2}-1)(b_k^{\dag}b_k+c_kc_k^{\dag}+b_{-k}c_{k}+c_{-k}^{\dag}b_{k}^{\dag})\\
\nn &+&
 \frac{\lambda a^3v^2cos^2(\rho)sin^2(\theta)}{4\omega_\pi}
(b_k b_{-k}+b_{k}c^{\dag}_k
 +c^{\dag}_k b_{k}+c_k c_{-k}+c^{\dag}_k
 c^{\dag}_{-k}+c_kb^{\dag}_k+b^{\dag}_{k}c_{k}+b^{\dag}_{k}b^{\dag}_{-k}))\\
 \nn
 &+& \frac{\lambda
a^3v^2cos(\rho)sin(\rho)sin^2(\theta)}{2\omega_\pi}
\lrb b_k a_{-k}+b_{k}a^{\dag}_k+c^{\dag}_k a_{k}+c_k
a_{-k}+c^{\dag}_k
a^{\dag}_{-k}+c_ka^{\dag}_k+b^{\dag}_{k}a_{k}+b^{\dag}_{k}a^{\dag}_{-k}\rrb\\
\nn
 &+ &\frac{\lambda
a^3v^2sin(\rho)sin(\theta)cos(\theta)}{\sqrt{\omega_\pi\omega_{\Sigma}}}\lrb
 d_k a_{-k}+d_{k}a^{\dag}_k+d^{\dag}_k a_{k}+d^{\dag}_k
 a^{\dag}_{-k}\rrb
 \nn \\
&+& \frac{\lambda
a^3v^2cos(\rho)sin(\theta)cos(\theta)}{\sqrt{\omega_{\pi}\omega_{\Sigma}}}
 \lrb
 b_k d_{-k}+b_{k}d^{\dag}_k+c^{\dag}_k d_{k}+c_k d_{-k}+c^{\dag}_k d^{\dag}_{-k}+c_kd^{\dag}_k+b^{\dag}_{k}d_{k}+b^{\dag}_{k}d^{\dag}_{-k}\rrb
 \}\eea
 where \be
\frac{\omega^{2}_{\pi}(k)}{a^{6}}\equiv\frac{\omega^{2}_{\pi_0}(k)}{a^{6}}=\frac{\omega^{2}_{\pi_\pm}(k)}{a^{6}}=(m_{\pi}^{2}+\frac{\veck^{2}}{a^{2}}):\:
\frac{\omega^{2}_{\Sigma}(k)}{a^{6}}=(m_{\Sigma}^{2}+\frac{\veck^{2}}{a^{2}})\ee
and
\bea
\frac{\Omega^2_{\pi_0}-\omega^2_{\pi_0}}{a^6}&=&\lambda[(<\Phi^2>-v^2)+2v_3^2]
\nn \\ \frac{\Omega^2_{\pi
{\pm}}-\omega^2_{\pi_{\pm}}}{a^6}&=&\lambda[(<\Phi^2>-v^2)+2v_+v_-]
\nn
\\
\frac{\Omega^2_{\Sigma}-\omega^2_{\Sigma}}{a^6}&=&\lambda[(<\Phi^2>-v^2)+2\sigma^2]
\eea

The background field  in the disoriented phase is given by :
\begin{equation}
\left(\begin{array}{c}v_+\\ v_-\\ v_3=v\\
\sigma\end{array}\right)\equiv <\Phi>,
\end{equation}
 In keeping with the disorientation of the vacuum on a three sphere we   parametrize the background field  through three angles:\be
<\Phi>=\lrb \ba{c}
vCos(\rho)Sin(\theta)Sin(\alpha)\\vCos(\rho)Sin(\theta)Cos(\alpha)\\vSin(\rho)Sin(\theta)\\vCos(\theta)\ea
\rrb.\ee
The dynamical evolution of  a system governed by this Hamiltonian
is considered  from  a  non equilibrium state state of restored
symmetry $<\Phi>=0$ at high temperatures to   the equilibrium
state of broken symmetry  $<\Phi>=v$ due to  the  subsequent
expansion and cooling of the QGP.   The time evolution can be
modeled through a time dependence of the parameter $<\Phi>(t)$,
which depends on the way in which the system relaxes to
equilibrium. In addition there is a time dependence of the
expansion factor $a(t)$ which allows for the rate of the expansion
of the plasma.  The following scenarios are possible:\\ {\bf A}. A
sudden quench from a state of restored symmetry to a state of
broken symmetry in a rapidly cooling expanding plasma, the
configuration of the  field  lags behind the expansion of the
plasma.If $\tau$ is the time the state spends in the symmetry
restored phase, a quench can be modeled by assuming that for
$0\let\le\tau$ the vacuum expectation value $<\Phi^2>=0$ and for
$t
>\frac{\tau}{2}$ the vacuum relaxes to its value
$<\Phi^2>=<v^2>$. In
 such a case the dynamical evolution of  $<\Phi>$ could be modeled by
 $
<\Phi >(t)=v \Theta[(t-\tau)]$
In addition in  a realistic scenario the expansion coefficient
$a(t)$ must also be a function of time and has been modeled
appropriately in ref [1]. For a sudden quench scenario the
Hamiltonian given in equation 1 diagonalizes  to\be H=\int
\momvol\frac{1}{2a^3}\{
\Omega_{\pi}(k,t)\{(A^{\dag}_{k} A_{k} +
\frac{1}{2}) +( C^{\dag}_{k} C_{k} + B^{\dag}_{k}
B_{k}+1)\}+\Omega_{\Sigma}(k,t)(D^{\dag}_{k}D_{k}+\frac{1}{2})\}.
\ee
by a unitary matrix given by a squeezed transformation \be
U(r,t)=e^{\int \momvol
r(k,t)\{(a^{\dag}_{k}a^{\dag}_{-k}-a_{k}a_{-k})+(d^{\dag}_{k}d^{\dag}_{-k}-d_{k}d_{-k})+(c_{k}b_{-k}+b_{k}c_{-k})-(c^{\dag}_{k}b^{\dag}_{-k}+b^{\dag}_{k}c^{\dag}_{-k})\}}
\ee
where $r_k$ is the squeezing parameter related to the physical
variables $\Omega_{\pi}(k,t)$ and $\omega_{\pi}(k)$ through \be
Tanh(2r_k)= \frac{(\frac{\Omega_k(t)}{\omega_\pi})^2-1}
 {(\frac{\Omega_k(t)}{\omega_\pi})^2+1} \ee The quench consists
of a rapid change of frequency of the time dependent harmonic
oscillator from $\Omega_\pi(k,t))$ to $\omega_\pi(k)$ though the
time evolution of $<\Phi>(t) $ and $a(t)$. It can be shown that
for a sudden frequency jump, such as that of a quench, there is
substantial squeezing, resulting in an enhancement of low momentum
modes signalling a DCC. The details of the evolution of the
Hamiltonian of a system undergoing a sudden quench (enhanced
squeezing) also show up in the difference in multiplicity
distributions of the charged and neutral pions illustrated in
Figure 1.
\begin{figure}[htbp]
\begin{center}
\epsfxsize=6cm\epsfbox{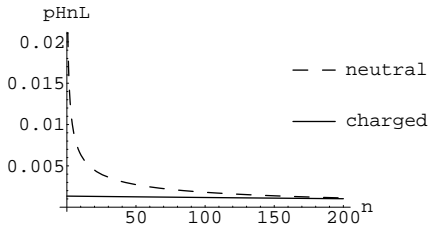}\end{center} \caption{Shows the
variation of $P_{n0}$ (solid line)and $P_{nc}$(dashed line) with
$n$ for the {\bf quenched} limit ($r_0=4$)}
\end{figure}

{\bf B}. Another non-equilibrium  situation that can arise  is
one where the  system can go through a metastable disordered
vacuum given by eqn[6] and then relax by quantum fluctuations to
an equilibrium configuration. Here  $\theta$  and $\rho$ measure
the degree of disorientation of the condensate in isospin space (
for charge conservation $\alpha=\frac{\pi}{4}$) . The
disorientation can be in the neutral sector $\rho=\frac{\pi}{2}$
in which case the angle $\theta $ mixes the neutral pion and sigma
field resulting in oscillations and enhancement of neutral pions
over the charged pions, we will call this case 1 . The
disorientation can be in the charged sector
$\theta=\frac{\pi}{2}$, where the angle $\rho$ mixes the charged
pions and the neutral pions resulting in charged oscillations and
enhancement of charged pions over the neutral pions we will call
this case 2. As an illustration of these effects we examine case 1
briefly. All the details of case 1 and 2 will be given in an
expanded later communication [2]. For case 1 the examination of
the Hamiltonian ( obtained by putting $\rho=\frac{\pi}{2}$ ) shows
a non-zero mixing term coming from the $\pi_{0}$-$\Sigma$ sector.
The misalignment of the vacuum through an angle $\theta$ induces a
mixing of the two fields. The mixed fields are
 \be
\bmat\Bk\\\Ck\emat=\bmat
Cos(\theta)\;\;\;Sin(\theta)\\-Sin(\theta)\;\;\;Cos(\theta)\emat\bmat\ak\\\dk\emat\ee
The diagonalization procedure for this case involves two squeezing
transformations of the mixed fields
 \bea
 \Fk&=&\mu A_{\theta(k)}+\nu A^{\dag}_{\theta(-k)} \nn\\
  \Gk&=&\rho D_{\theta(k)}+\sigma
D^{\dag}_{\theta(-k)}.  \eea The diagonalized form is
\be H_{\pi/2}=\int \momvol \{ \frac{\Omega_{\pi_\pm}}{a^3}(
C^{\dag}_{k} C_{k} +
B^{\dag}_{k}B_{k}+1)+\frac{\sqrt{\Omega_{\pi}\Omega_{\Sigma}}}{4
a^3}\lrb(\Fdagk\Fk+\frac{1}{2})+(\Gdagk\Gk+\frac{1}{2})\rrb\}\ee
During time evolution again we have a time dependent frequency for
the state of mixed neutral pions and sigma   which changes from
$\frac{\sqrt{\Omega_{\pi}(k,t)\Omega_{\Sigma}(k,t)}}{4 a^3}$
to$\frac{\sqrt{\omega_{\pi}(k,t)\omega_{\Sigma}(k,t)}}{4 a^3}$
this time through the time evolution of $\theta(t) $ from the
disoriented value to 0 and $a(t)$. There is again an enhancement
of neutral pions as a result of the DCC formation and furthermore
a  mixing in the neutral sector . In addition there are
oscillations in the number of neutral pions as the disorientation
$\theta$ changes with time shown  in Figure 2.
\begin{figure}[htbp]
\begin{center}
\epsfxsize=6cm\epsfbox{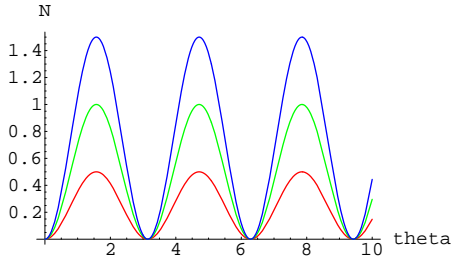}\end{center} \caption{Shows the
oscillations in the number of neutral pions as a function of the
disorientation $\theta$ for the value of the squeezing parameters
$"red=\nu=.5,green=\nu=1,blue=\nu=1.5$}
\end{figure}

A case of particular interest , in view of latest preliminary
experimental results  is case 2 i.e. $\theta=\frac{\pi}{2}$
because there is a mixing of charged and neutral fields due to the
disorientation. This gives rise to interesting charge-neutral
fluctuations  which can be measured in  a DCC detection
experiment[3]. A full discussion of the detailed theory  is given
in [2].

To conclude we have modeled  the evolution of the disoriented
chiral condensate  through both a sudden quench and with a
transition through a metastable state with arbitrary
disorientation and have shown that the total multiplicity
distributions of charged and neutral pions functions are dramatic
characteristic signals for the DCC  and are related directly to
the way in which the DCC forms. These are unambiguous, therefore
they must be examined thoroughly in searches for the DCC [4]. We
are encouraged by the first preliminary data of references [3] and
[4], from the analysis of the WA98 experiment at the CERN SPS in
which some events show an excess of photons (neutral pion excess)
within the overlap region of charged and photon multiplicity
detectors, the preliminary results given in ref [4] are shown in
figure 3.\begin{figure}[htbp]
\begin{center}
\epsfxsize=6cm\epsfbox{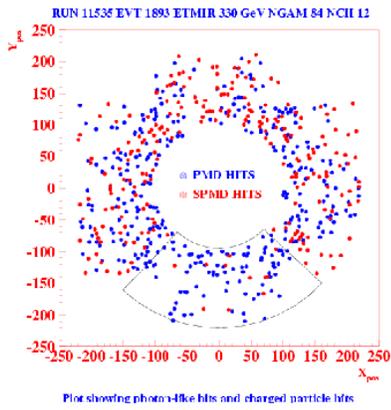}\end{center} \caption{Shows the
photon hits and charged particle hits in WA98 from ref. 4}
\end{figure} In view of the results presented in this section, we
hope that the future will bring more exotic events with charge
excess and neutral and charged multiplicity oscillations.
\newpage

{\bf References}
\begin{itemize}
\item[1.] B. Bambah and C. Mukku, Phys. Rev. D. {\bf70 }(3) 0340001
(2004).\\B.A. Bambah and C.Mukku, Annals of Physics 314,
34,(2004)\item[2.] B. A. Bambah , C. Mukku and K.V.S Shiv
Chaitanya {\it Dynamics of the Dcc with Arbitrary Disorientation
in Isospin Space}( In preparation).
\item[3.] T. Nayak ,Pramana 57 ,285,(2001).
\item[4.] M.M. Aggarwal et al., e-print archive:nucl-ex/0012004.\\ Madan M
Aggarwal, Pramana {\bf 60}, No.5,   987 , (2003)
\end{itemize}

\end{document}